\documentclass[journal,letterpaper]{IEEEtran}
\usepackage{amsmath,amssymb,amsfonts}
\usepackage{array}
\usepackage{graphicx}
\usepackage{cite}
\usepackage{epstopdf}
\usepackage{subfigure}
\usepackage{subfigure}
\usepackage{color}
\usepackage{multicol}
\def\BibTeX{{\rm B\kern-.05em{\sc i\kern-.025em b}\kern-.08em
    T\kern-.1667em\lower.7ex\hbox{E}\kern-.125emX}}

\begin{document}

\title{Photo-caloritronics in the THz regime: Photon-assisted thermopower generators}
\author{Parijat~Sengupta, Saptarshi~Das, and Junxia~Shi 
\thanks{Parijat Sengupta and Junxia Shi are with the Department of Electrical and Computer Engineering, University of Illinois, Chicago, IL, 60607, USA.}
\thanks{Saptarshi Das is with the Department of Engineering Science and Mechanics, Pennsylvania State University, State College, PA, 16802, USA.}}

\maketitle

\begin{abstract}
The generation of thermopower in a miniaturized device modeled as a channel connected to reservoirs/contacts maintained at different temperatures is studied in this work. The left reservoir is also coupled to a periodic THz driving source leading to a rearrangement of its internal energy levels. The modified levels are determined using Floquet theory that governs the dynamics of a periodic Hamiltonian. The starting point for the presented thermopower calculations (mirrored in the related thermal gradient controlled current) is the Landauer formula rooted in the transmission formalism. Primarily, we show that while thermally activated electrons can be pumped from the hot reservoir into the cold side as a straightforward manifestation of the Seebeck effect through a difference in their respective Fermi levels, the quantum of charge flow increases in presence of the periodic perturbation. We explain this phenomenon by noting how the periodic driving makes available a greater number of states in the left reservoir that are able to inject electrons into the channel. The calculations also uncover a useful feature whereby the strength of such a thermally-pumped current is amenable through a joint control of the amplitude and frequency of the signal, offering an additional experimentally-adjustable tool to regulate their flow. The calculated results are shown for two classes of materials defined by prototypical linear and quadratic dispersion, with the former capable of furnishing a larger current by virtue of higher available density-of-modes. We close by pointing out possible improvements to the calculation by accounting for dissipative effects in the channel. 
\end{abstract}


\vspace{0.2cm}
\section{Introduction}
\vspace{0.2cm}
The control and modulation of thermally-driven electric currents constitute an essential step in the energy harvesting paradigm~\cite{snyder2009thermoelectric,sothmann2014thermoelectric,benenti2017fundamental} that pivots around the interplay between charge and heat flow. This traditional viewpoint, however, has been since revised and the coupling of heat and charge can be possibly mediated with electron's other accessible degrees-of-freedom (DOF), for example, spin~\cite{bauer2012spin,boona2014spin}. This correlation between transfer of heat and the spin DOF, like all other such conjoined processes, is closely tied to the theme of the ease of manipulation of the intrinsic microscopic character at the nanometer scale, which, manifesting under the right conditions can lead to a significantly altered heat response. Such manipulations have collectively given rise to a broad class of metamaterials~\cite{engheta2006metamaterials,padilla2007electrically} often touted for their utility in advancing nanoscale phenomena including the construction of miniaturized heat engines. These modifications, in principle, can be also brought to realization through the interaction of light with matter, amply demonstrated in a wide array of optical effects~\cite{herek2002quantum,rakheja2016tuning} when an illuminated beam or photon source is absorbed by the charge carriers of a crystal. A particular effect in this regard involves the observation of a cooperative phenomenon between the time-periodic electric field of a light/photon source and the electronic setup of a material, resulting in the latter exhibiting an altered energy manifold with additional quasi-states. The introduction of such quasi-states isn't simply a mathematical artifact within the purview of the Floquet theory~\cite{dittrich1998quantum,tannor2007introduction} that deals with dynamics of systems perturbed by time-periodic signals, but has far-reaching and non-trivial consequences; for instance, it has been shown that such external periodic inducements can bring about the emergence of superconductivity~\cite{benito2014floquet,raines2015enhancement} in semiconductors, the onset of topological phase transition in graphene~\cite{delplace2013merging,lindner2011floquet}, and the presence of a pseudo-thermal state in the charge density wave phase of quantum chains~\cite{kennes2018floquet}. In this work, we seek to employ such Floquet/quasi-states in the quantitative assessment of the quantum of thermal current flowing in an appropriately conceived thermopower generator.

The value of such calculations that estimate a thermal current lie in the current trend to treat thermoelectric systems so as to construct the quantum counterparts of classical heat engines and refrigerators, but with greater efficiency than achieved heretofore when solely governed by thermodynamics. Moreover, classical devices usually exemplified by the archetypal Carnot engine rely on pistons and plungers and a combination of moving parts rendering them essentially unusable in an ever-shrinking feature size of electronic architecture. It is therefore of some merit to consider alternative techniques that unifies elements of quantum theory with thermodynamics - the field of quantum thermodynamics~\cite{kosloff2013quantum}. We begin by first describing the elements of the suggested thermopower generator operated in conjunction with a periodic \textit{ac}-source, the choice of thermoelectric materials, and a summary of results. A more detailed account is contained thereafter starting with the relevant theory in Section~\ref{s2} wherein a complete analysis of the model structure is carried out employing the Floquet theory. The analytic results in the first half serve as a starting point for subsequent current expressions followed by its numerical estimation (Section~\ref{s3}) for a set of material and experimental parameters. We conclude (Section~\ref{s4}) by pointing out a set of possible refinements that may bring the theory outcomes in much closer alignment with experimental measurements. 

\begin{figure*}
\center
\includegraphics[width=0.7\textwidth]{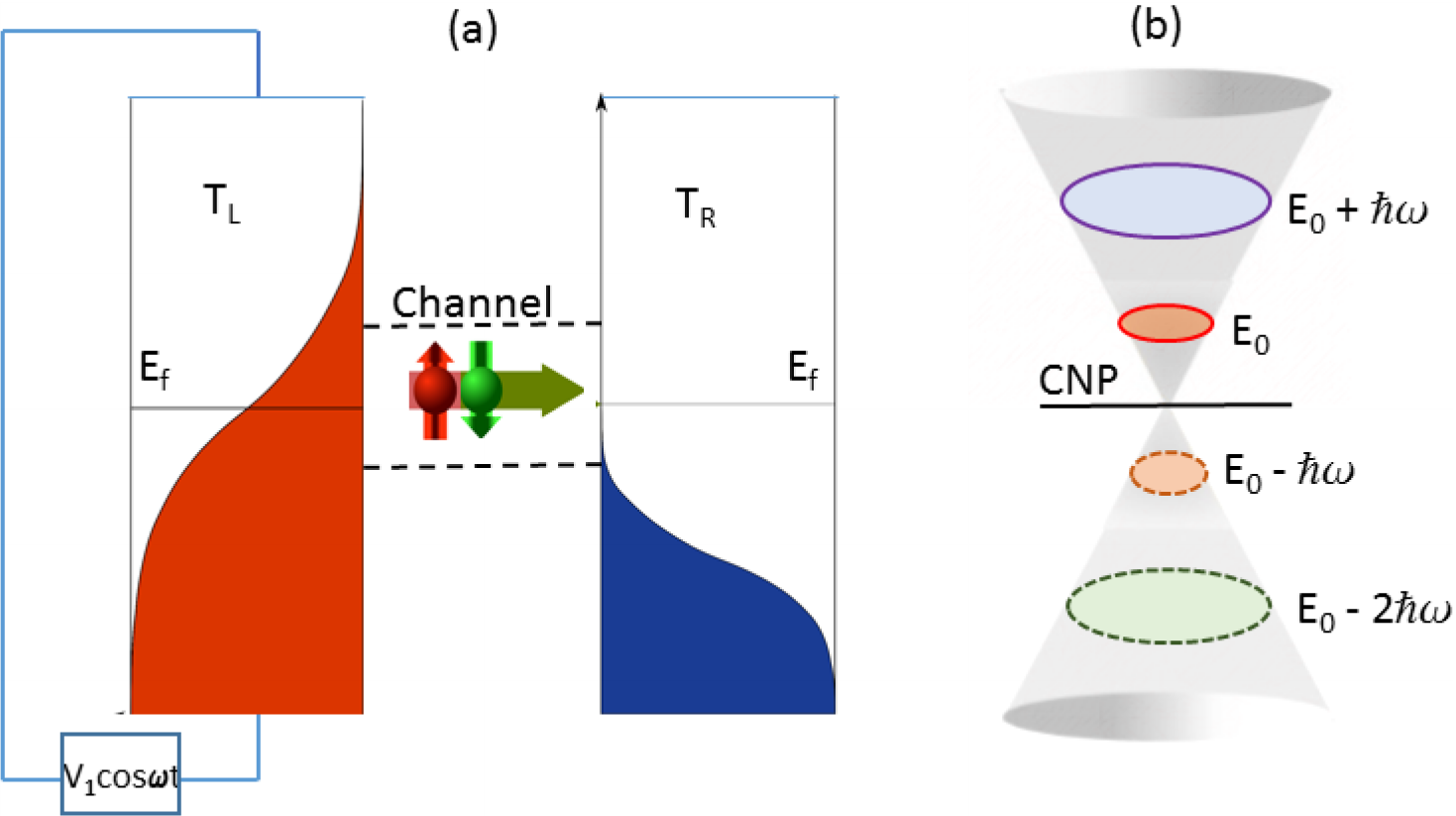}
\caption{The left panel (a) shows the schematic of a channel material with two reservoirs/contacts affixed to its ends and maintained at different temperatures. The left (right) reservoir is at temperature $ T_{L} \left(T_{R}\right) $ and $ T_{L} > T_{R} $. The temperature difference $\left(\Delta T\right)$ between the reservoirs sets up a thermal current solely reliant on the corresponding Fermi level $\left(E_{f}\right)$ difference. This current contains both sets of spin carriers marked by vertical arrows are assumed to flow equally. In addition to $ \Delta T $, the left reservoir is fitted with a time-dependent periodic drive $\left(V_{1}\cos\omega t\right) $ that introduces `extra' energy levels in multiples of $ \hbar\omega $ (the drive energy). The thermal current via the photon-induced states is adjustable by tuning the external periodic perturbation. The figure on the right (b) illustrates the new energy side-bands created by the time-periodic drive in a graphene-like Dirac material. The Dirac crossing or the charge neutral point (CNP) is the meeting of the funnel-shaped conduction and valence bands. The photon-created states have been shown using a base energy level of $ E_{0} $. Note that such bands are, however, not limited to Dirac materials and have been observed in systems with conventional parabolic dispersion. Further, the channel material in the analysis that follows considers both parabolic and linear bands and explicates the quantitative difference in the quantum of photon-modulated thermal current for the two cases. A discussion later in the text (see Fig.~\ref{bescoff}) works out the number of such bands that must be included to accurately portray their role in the current setup.}
\label{scfig1}
\vspace{-0.3cm}
\end{figure*}

The proposed scheme illustrated in Fig.~\ref{scfig1} is a double-T structure consisting of two large pads (henceforth referred to as reservoirs) joined by a narrow constriction. A time-dependent drive $\left(V\cos\omega t\right) $ which mimics the electric field of light is applied to the left pad via a top metal gate. For an easier analysis that does not mask the central theme of this paper, we consider the system to be homogeneous (carved out of a single material); quantitative implications when this assumption is relaxed are touched upon later. We further allow the possibility wherein the structure in Fig.1 could be designed with materials that identify as of the Dirac-type (graphene-like possessing linear dispersion) or the conventional GaAs-like parabolic description. Briefly, we find that the generated thermopower records an uptick in presence of an external \textit{ac}-source and can be further adjusted by tuning the drive strength, the ratio of amplitude to frequency of the periodic signal. The thermopower, for a given signal frequency, can be scaled up (down) through decrements (increments) of the amplitude - a result we explain by studying the respective `weight' of each contributing harmonic represented by a Bessel's function expansion. Lastly, the thermopower for the suggsted configuration is shown to be always higher for linear materials vis-\'a-vis their parabolic counterparts, the presence of an external \textit{ac}-source does not alter this conclusion. 

As a final note, the frequency of the \textit{ac}-signal or the periodic perturbation is assumed to lie in the THz $\left(10^{12}\right)$ spectral range defined as the region from 0.1 to 30 THz. This frequency range of late has been used to identify different physical mechanisms~\cite{dexheimer2007terahertz,ganichev2006intense} in diverse material systems and utilized in laboratory demonstrations of detection and sensing elements, notably thermal sensors and photo-acoustic conversion detectors. The practical utility in THz microelectronic technology aside, a more intriguing feature is their characteristic interaction with matter~\cite{borak2005toward}, an instance of which is presented here. The rise of THz applications is also spurred by an easy availabilty of an array of laser (ultra-femtosecond lasers) modes and oscillating circuits~\cite{lewis2014review} whose frequencies fall in the desired spectrum offering a wide choice of appropriate generation sources. Recently, more efficient THz sources offered as commercial products make use of quantum cascade lasers (QCL)~\cite{williams2007terahertz}. 

\section{Theoretical formalism}
\label{s2}
For a quantitative analysis we begin by setting up the Hamiltonian under an external periodic drive that introduces the Floquet states in a generic two dimensional slab of a parabolic material. In its simplest form, the Hamiltonian can be written as
\begin{equation}
i\hbar\dfrac{\partial}{\partial t}\Psi\left(r,t\right) = \left[-\dfrac{\hbar^{2}}{2m}\nabla^{2} + V_{0} + V_{1}cos\omega t\right]\Psi\left(r,t\right).
\label{schpdr} 
\end{equation}
In Eq.~\ref{schpdr}, the Hamiltonian can be split into a time independent and dependent part. The first two terms are time independent $\left(H_{0}\right)$ while the time dependent part is the driving term, $ V_{1}\cos\omega t $. Performing a separation of the spatial and temporal variables, the wave function can be written as $ \Psi\left(r,t\right) = \phi_{1}\left(r\right)\phi_{2}\left(t\right) $ and noting that $ H_{0}\phi_{1}\left(r\right) = E\phi_{1}\left(r\right) $, we obtain the following differential equation for $ \phi\left(t\right) $
\begin{equation}
i\hbar\dfrac{\partial}{\partial t}\phi_{2}\left(t\right) = \left[E + V_{1}\cos\omega t\right]\phi_{2}\left(t\right).
\label{schtpd}
\end{equation}
The solution for the temporal part of the wave function from Eq.~\ref{schtpd} is easily written as 
\begin{equation}
\phi_{2}\left(t\right) =  \exp\left[-i/\hbar\left(Et + \dfrac{V_{1}}{\omega}\sin\omega t\right)\right]. 
\label{tempwf}
\end{equation}
Expanding the sine term in a Fourier expansion, the solution can be more compactly written as
\begin{equation}
\phi_{2}\left(t\right) = \exp\left(-iEt/\hbar\right)\sum_{-\infty}^{\infty}a_{n}\exp\left(-in\omega t\right).
\label{tempsol}
\end{equation}
Employing the standard procedure to determine Fourier coefficients, the following relation can be written
\begin{equation}
a_{n} = \dfrac{1}{2\pi}\int_{-\pi}^{\pi}\exp\left[i\left(n\omega t - \dfrac{V_{1}}{\hbar\omega}\sin\omega t\right)\right].
\label{focoeff}
\end{equation}
The complete time-dependent part of the wave function is therefore
\begin{equation}
\phi_{2}\left(t\right) = \sum_{-\infty}^{\infty}\biggl[J_{n}\left(x\right)\exp\left[-i\left(E + n\hbar\omega\right)t/\hbar\right]\biggr],
\label{bessel}
\end{equation}
where $ J_{n}\left(x\right) $ is the $ n^{th} $-order Bessel's function of the first kind and $ x = V_{1}/\hbar\omega $. Note that this particular form of solution is guided by the equation defining the coefficient $ a_{n} $ - the integral form of $ J_{n} $. The complete wave function is therefore $ \Psi\left(r,t\right) = \phi_{1}\left(r\right)\sum\limits_{-\infty}^{\infty}\biggl[J_{n}\left(x\right)\exp\left[-i\left(E + n\hbar\omega\right)t/\hbar\right]\biggr] $. An interpretation~\cite{platero2004photon} of this solution is to view it as a series of energies $\left(E + n\hbar\omega\right)$ marked by an associated finite density of states with $ J_{n}^{2}\left(x\right) $ probability of occurrence. The processes of absorption and emission correspond to $ n > 0 $ and $ n < 0 $, respectively. As a useful remark, note that the aforementioned set of steps correspond to the well-known Tien-Gordon model~\cite{tien1963multiphoton}, which was first carried out for a microwave-driven superconductor-insulator-superconductor tunnel junction. This model is a subset of the more complete Floquet theory of treating time-periodic differential equations.

\subsection{Temperature and frequency modulated current}
\begin{figure}[t!]
\centering
\includegraphics[scale=0.6]{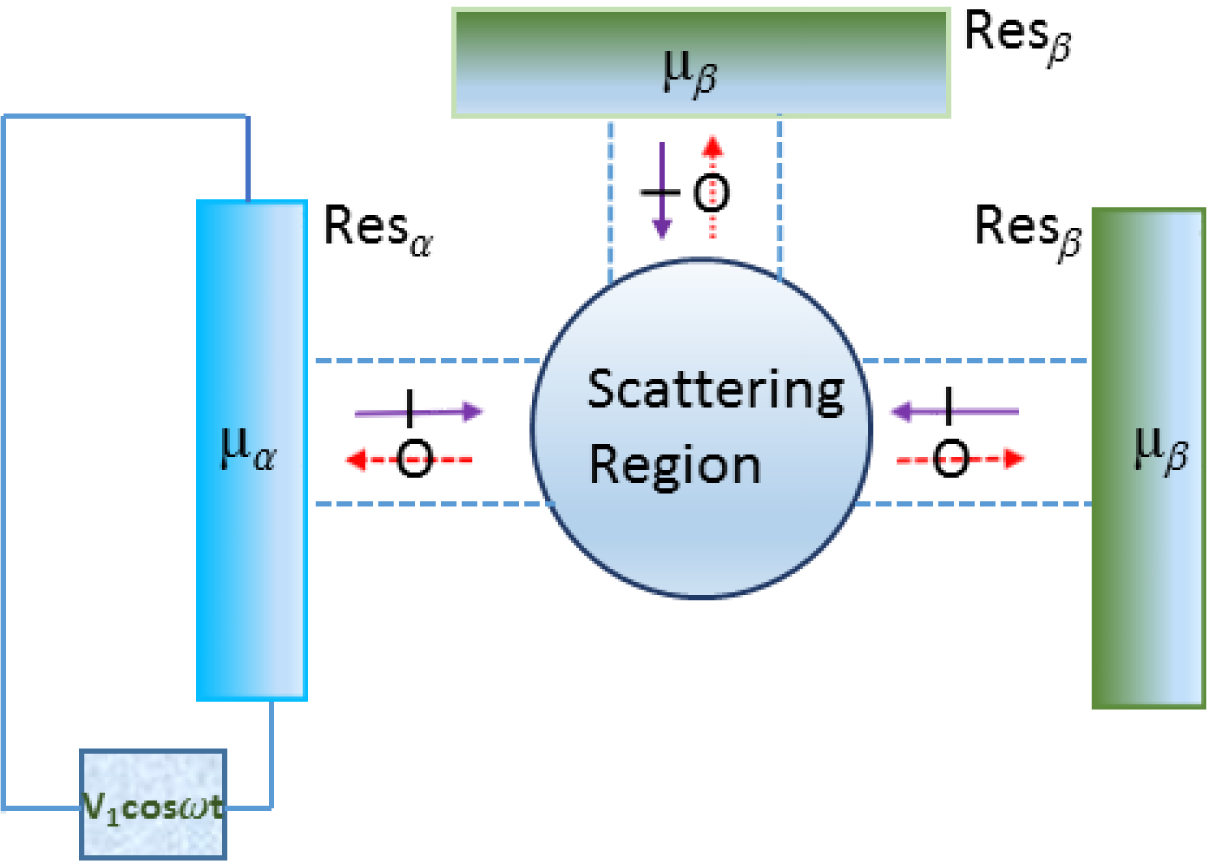}
\caption{The illustration shows a generic central scattering region (the constricted channel of Fig.~\ref{scfig1} in the present case connected to a number of reservoirs/contacts, which are assumed to extend to infinity. In this multi-terminal/lead arrangement, the scattering matrix $\left(S_{\alpha\beta}\left(E\right)\right)$ that typifies the central region (see text for an elucidation about the nature of this matrix) connects the entering particles (incoming waves marked by solid lines with an `I') from all the leads collectively identified as Res$_{\beta}$ and their draining (particles leaving via an outgoing wave denoted by broken lines and an `O') into the reservoir, Res$_{\alpha}$. In the completely general case, there exists a finite reflection $\left(r\right)$ and transmission $\left(t\right)$ probability at the junction between the reservoirs and the channel; here, however, we assume a reflection-less junction or expressed as, $ r = 0 $ and $ t = 1 $. Additionally, while carriers in each reservoir characterized by their respective Fermi level $\left(\mu\right)$ are in equilibrium, the possibility of $ \mu_{\alpha} \neq \mu_{\beta} $ gives rise to a non-equilibrium situation and a current flow. In this report, we simulate a non-parity between the Fermi levels of reservoirs/contacts by virtue of a temperature gradient. The time periodic external drive that builds a cascade of states in one of the reservoirs and useful for purpose of modulation of the temperature gradient controlled current is in the form of a sinusoidal-type $\left(V_{1}\cos\omega t\right)$ perturbation.}
\label{scattpic}
\vspace{-0.3cm}
\end{figure}

We adopt the $ S $-matrix formalism~\cite{buttiker1992scattering,datta1997electronic} and write the per-spin current operator following Ref.~\citenum{platero2004photon} in a lead $ \alpha $ as
\begin{equation}
\begin{aligned}
\hat{I}_{\alpha}\left(t\right) &= \dfrac{e}{\hbar}\int dE\int dE^{'}\exp\left(\dfrac{it}{\hbar}\left(E - E^{'}\right)\right) \\
&\times \biggl[\hat{a}_{\alpha m}^{\dagger}\left(E\right)\hat{a}_{\alpha m}\left(E^{'}\right) - \hat{b}_{\alpha m}^{\dagger}\left(E\right)\hat{b}_{\alpha m}\left(E^{'}\right)\biggr].
\label{cueqn}
\end{aligned}
\end{equation}
The creation operator is $ \hat{a}_{\alpha m}\left(E\right) $ while the corresponding annihilation operator is expressed as $ \hat{a}_{\alpha m}^{\dagger}\left(E\right) $. Here, the subscript $ m $ denotes a specific mode contained in a lead $ \alpha $. In short hand notation, $ m \in \alpha $. The two sets of creation and annihilation operators in Eq.~\ref{cueqn} are connected via the scattering matrices. They are expressed as:
\begin{equation}
\begin{aligned}
\hat{b}_{\alpha n}\left(E\right) &= \sum_{m,\beta}s_{nm}^{\alpha \beta}\hat{a}_{\beta m}\left(E\right),\\
\hat{b}_{\alpha n}^{\dagger}\left(E\right) &= \sum_{m,\beta}\left(s_{nm}^{\alpha \beta}\right)^{\dagger}\hat{a}_{\beta m}\left(E\right).
\label{scatop}
\end{aligned}
\end{equation}
This description of scattering matrices (Eq.~\ref{scatop}) conveys information about the outgoing wave via mode $ n $ in lead $ \alpha $ from the scattering region that receives incoming waves from all available modes $\left(m\right)$ in a certain lead $\left(\beta\right)$ of a multi-terminal structure. A pictorial depiction of this situation is shown in Fig.~\ref{scattpic}. 
Substituting for the creation and annihilation operators representing the outgoing state (Eq.~\ref{scatop}) in the current equation (Eq.~\ref{cueqn}), we have the form
\begin{equation}
\begin{aligned}
\hat{I}_{\alpha}\left(t\right) &= \dfrac{e}{\hbar}\sum_{m \beta}\int dE\int dE^{'}\exp\left(\dfrac{it}{\hbar}\left(E - E^{'}\right)\right) \\
&\times \biggl[\hat{a}_{\beta m}^{\dagger}\left(E\right)A_{\beta \gamma}\hat{a}_{\gamma n}\left(E^{'}\right)\biggr].
\label{cueqn1}
\end{aligned}
\end{equation}
The operator $ A_{\beta \gamma}\left(\alpha, E, E^{'}\right) $ in its most general form is:
\begin{equation}
A_{\beta \gamma}\left(\alpha, E, E^{'}\right) = \delta_{\alpha \beta}\delta_{\alpha \gamma}\mathbf{1}_{\alpha} - s^{\dagger}_{\alpha \beta}s_{\alpha \gamma}.
\label{aopt}
\end{equation}
For brevity, the energy dependence and explicit reference to lead $ \alpha $ in case of the $ A $ operator is not shown in Eq.~\ref{cueqn1}. The notation $ \beta $ and $ \gamma $, as before, identifies two distinct leads with modes $ m $ and $ n $, respectively. Since we are interested in the quantum statistical average current, it  is reasonable to set $ E = E^{'} $ and add the contribution of each lead separately (accounting for all current carrying modes located within it). In Eq.~\ref{aopt}, therefore, in addition to selecting a single energy, we also set $ \beta = \gamma $. Note that through the substitution $ E^{'} = E $, the exponential term goes to unity. This summation of all leads independently simply represents the trace of the $ A $ matrix while the combined creation and annihilation operators simply indicate the occupation number. Succinctly, it is expressed through the following relation
\begin{equation}
\hat{a}_{\beta}^{\dagger}\left(E\right)\hat{a}_{\gamma}\left(E\right) = \delta_{\beta \gamma}\sum_{n}J_{n}^{2}\left(x\right)f_{\beta}\left(E + n\hbar\omega\right).
\label{fermirel}
\end{equation}
In Eq.~\ref{fermirel}, the factor $ \sum_{n}J_{n}^{2}\left(x\right) $ is introduced by virtue of a modified creation and annihilation operator to account for the $ J_{l}\left(x\right) $ term in the wave function (Eq.~\ref{bessel}). We write it as $ \hat{a}_{\beta}\left(E\right) = \sum_{l}\hat{a}_{\beta}^{'}\left(E + l\hbar\omega\right)J_{l}\left(x\right) $. Carrying out the trace operation on the $ A $ matrix and inserting the result from Eq.~\ref{fermirel} in the generalized current expression (Eq.~\ref{cueqn1}), the final form for a two-terminal device sketched in Fig.~\ref{scfig1} is written as 
\begin{equation}
\begin{aligned}
\hat{I}_{\alpha}\left(t\right) &= \dfrac{e}{\hbar}\sum_{n}J_{n}^{2}\left(x\right)\int dE \mathcal{T}\left(E\right)\mathcal{M}\left(E\right)\\
&\times \left[f_{\alpha}\left(E + n\hbar\omega\right) - f_{\beta}\left(E\right)\right].
\label{finceqn}
\end{aligned}
\end{equation}
The full set of steps leading to Eq.~\ref{finceqn} can be found, for example, in Refs.~\cite{platero2004photon,heikkila2013physics}. To reiterate, in this two-terminal arrangement, $ \mathcal{T}\left(E\right) $ is the transmission probability summed over all modes $\left(\mathcal{M}\left(E\right)\right)$ from lead $ \beta $ to lead $ \alpha $. The expression in Eq.~\ref{finceqn} allows a quantitative determination of the quantum of current that flows between the two leads; however, we must first fix the transmission and enumerate the modes available.

As a first approximation, we set the transmission to unity assuming the target structure (see Fig.) to be homogeneous throughout. To estimate $ \mathcal{M}\left(E\right) $, as is standard practice, we assume periodic boundary conditions along the \textit{y}-axis such that the $ k $-channels are equi-spaced by $ 2\pi/W $.~Here, $ W $ is the width (the transverse span along the \textit{y}-axis) of the sample. Each unique $ k $-vector is a distinct mode and the number of such momentum vectors is determined from the inequality, $ -k_{f} < k_{y} < k_{f} $. The upper and lower bounds of the inequality are the momentum vectors that correspond to the Fermi energy, E$_{f}$. The approximate number of modes is therefore $ k_{f}W/\pi $. Bearing these in mind, Eq.~\ref{finceqn} can be recast as (dropping the operator notation for current)
\begin{equation}
\begin{aligned}
I_{\alpha} &= \dfrac{eW}{\pi h}\sum_{n}J_{n}^{2}\left(x\right)\int g\left(E\right)dE\int_{-\pi/2}^{\pi/2}d\theta \cos\theta \\
&\times \left[f_{\alpha}\left(E + n\hbar\omega\right) - f_{\beta}\left(E\right)\right]. 
\label{lbfcun}
\end{aligned}
\end{equation}
The function $ g\left(E\right) $ in Eq.~\ref{lbfcun} is the analytic representation of the $ k $-vector in energy space and evidently connected to the material dispersion. As noted above, we consider two distinct dispersion cases here: 1) Conventional parabolic (\textit{P}) expressed as $ E_{P} = \alpha k^{2} + \Delta $, where $ \alpha = \hbar^{2}/2m^{*} $ and 2) Linear Dirac (\textit{L}) which is $ E_{L} = \sqrt{\left(\beta k\right)^2 + \Delta^{2}} $. The effective mass is $ m^{*} $ and $ v_{f} = \beta/\hbar $ is the Fermi velocity. For completeness, a band gap, $ \Delta $, has been added to both dispersion forms. The function $ g\left(E\right) $ in Eq.~\ref{lbfcun} for the two cases (\textit{P} \& \textit{L}) can therefore be straightforwardly written as:
\begin{subequations}
\begin{equation}
g_{P}\left(E\right) = \sqrt{\dfrac{1}{\alpha}\left(E - \Delta\right)},
\label{mopar}
\end{equation}
and
\begin{equation}
g_{L}\left(E\right) = \sqrt{\dfrac{1}{\beta^{2}}\left(E^{2} - \Delta^{2}\right)}.
\label{molin}
\end{equation}
\end{subequations}
As a clarifying note, the sum of modes along the width $\left(W\right) $ is $\sum_{k_{f}W/\pi}\Delta k_{y} $ which in the continuum limit changes to $ \int_{-k_{f}} ^{k_{f}}dk_{y} $. As usual, the azimuthal angle is $ \theta $ while $ k_{x} = k\cos\theta $ and $ k_{y} = k\sin\theta $. Note that the limits of angular integration satisfies the span of a given $ k $-vector $\left(-k_{f} < k_{y} < k_{f} \right) $ or mode. Before we proceed to quantitative indicators of the presented formalism, a caveat must be included here: The Floquet theory predicts additional energy levels symmetrically located above and below (see, for example, such states pictorially drawn for graphene in Floquet mode in Fig.~\ref{scfig1}) $ n = 0 $, the base energy, which for our case is bottom of the conduction band. The symmetry property arises from the well-known property of Bessel functions~\cite{prosperetti2011advanced}: $ J_{-\vert n\vert}\left(x\right) = -1^{\vert n\vert}J_{\vert n \vert}\left(x\right)$. However, for a Fermi level $\left(\mu\right)$ placed at $ n = 0 $ mark, we may choose to ignore the $ n < 0 $ states, an approximation justified by noting that as such the difference in Fermi potentials (of the reservoirs/contacts) between filled states (the $ n < 0 $ states lie below $ \mu $) is vanishingly small.

\begin{figure*}[!ht]
\center
\includegraphics[width=0.6\textwidth]{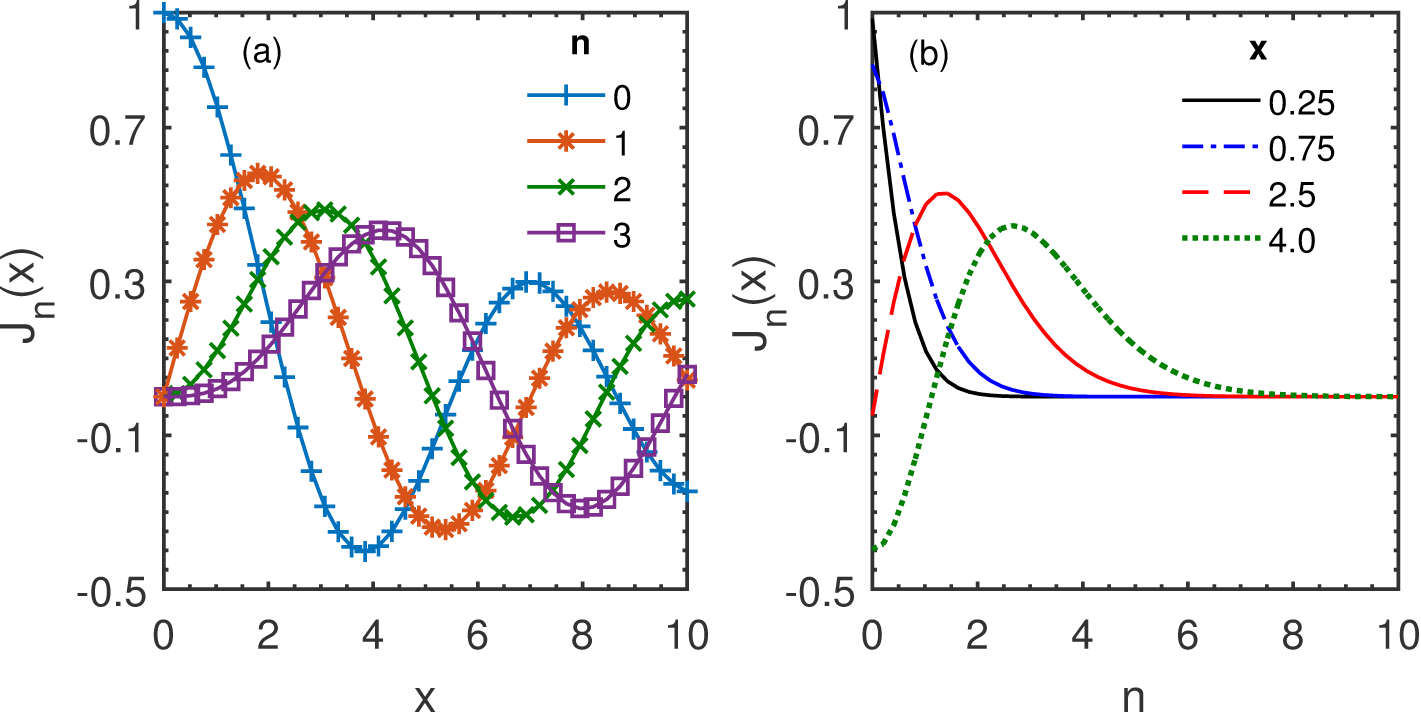}
\caption{The plots for the Bessel's function $ J_{n}\left(x\right) $ for order $ n = 0, 1, 2, 3 $ are shown on the left panel (a) for several values of $ x $. Clearly, the inverse factorial scaling `shrinks' the Bessel function profile for higher values of $ n $. Correspondingly, we also plot $ J_{n}\left(x\right) $ profiles (b) for different $ n $ values with a preset $ x $; the graphs again reflect a damping of the Bessel's function for $ n > 2 $. However, using the plot on the right panel as a guide, it can be easily seen that for higher values of $ x  = V_{1}/\hbar\omega $, which simply means a more robust perturbation, the Bessel's function still receives sizable contributions beyond the second order $\left(n = 2\right)$. The higher-order curves (or a greater $ n $) flatten out relatively farther into the \textit{x}-axis as compared to the case when $ x $ is less than unity. In the family of Bessel function curves, the overall statistical measure swings between the values of $ x $ and $ n $; a large $ x $, which in our case is incumbent upon the amplitude and frequency of the perturbation requires the inclusion of a greater number of photon-induced energy bands. Note that the probability of each additional band is $ J_{n}^{2}\left(x\right) $. Regardless, all calculations here are restricted up to the $ n = 2 $ order; implications of such a `hard' 
cutoff are noted in the closing section. The Bessel curves were plotted using the `besselj' function available in MATLAB.}
\label{bescoff}
\end{figure*}

\section{Numerical estimate of the thermal current}
\label{s3}
A quantitative demonstration of the photon-processes controlled thermal current involves summing the various energy levels $\left(\sum\limits_{m}\right)$ introduced via a periodic driving; the energy levels as outlined above have a definite probability tied to the $ n^{th} $-order Bessel's function $ \left(J_{n}\left(x\right)\right) $ of the first kind. In general $ n \rightarrow \infty $, however, a numerical solution must impose a cut-off to evaluate the sum. We carry out that procedure by first noting that the Bessel's function can be expressed as a power series of the form~\cite{prosperetti2011advanced}
\begin{equation}
J_{n}\left(x\right) = \sum\limits_{m = 0}^{\infty}\left(-1\right)^{m}\dfrac{\left(x\right)^{2m + n}}{2^{2m + n}m!\Gamma\left(m + n + 1\right)!}.
\label{besexp}
\end{equation}
For positive integral values of $ x = V_{1}/\hbar\omega $, the gamma function $\left(\Gamma\right)$ is given by $ \Gamma\left(n + 1\right) = n! $. Plotting the Bessel's function using Eq.~\ref{besexp}, it is clearly evident from Fig. that it damps for large $ n $ values on account of the increasing factorial in the denominator. The damping pattern is true regardless of the magnitude of the argument $ x $. For the cases considered here, we set the condition $ 0.25 \leq x \leq 4.0 $ which reflects a weak-to-moderate drive strength such that terminating the Bessel's power series (Eq.~\ref{besexp}) at $ n = 2 $ is a reasonable choice. For values of $ n $ beyond the set cut-off, the probability strength $ J_{n}^{2}\left(x\right) $ is weak (see Bessel plots, Fig.~\ref{bescoff}) and safely discarded. An illustration of the occurrence of additional side bands is also shown in Fig.~\ref{scfig1}. As a final remark, observe that when $ x $ is close to zero (for instance, $ x = 0.25 $ in Fig.~\ref{bescoff}), the contribution of the side bands tail off insofar as that the main energy band becomes the primary contributor to any microscopic interaction. For significant participation of the Floquet bands, $ x $, must be considerably larger than unity, a condition achieved through higher amplitude $\left(V_{1}\right)$ values of the periodic perturbation and a reasonably low frequency $\left(\omega\right)$. 

\subsection{Zero-illumination thermopower}
We start by carrying out a formal comparison between the thermal current that flows in linear and parabolic materials under a temperature gradient (see Fig.~\ref{scfig1}. For purpose of numerical estimation, let us first assign values to a few essential parameters that enter the current expression in Eq.~\ref{lbfcun}. For the parabolic material, the effective mass is set to $ 0.0565m_{0} $, where $ m_{0} = 9.1 \times 10^{-31} kg $ is the free electron mass. In practice, this corresponds to the transverse mass of conduction electrons located at the $ L $-valley of a $ 6.0\,nm $ wide lead chalcogenide PbTe thin film, which is modeled as a quantum well. The direct $ L $-valley band gap $\left(\Delta_{P}\right)$ at such temperatures for PbTe can be set to $ 0.19 eV $. The effective mass in this case was calculated using a standard $ 4 \times 4 $ k.p Hamiltonian for lead chalcogenides~\cite{kang1997electronic}; for a full analysis, relevant band structure parameters, associated computational details, and the overall matrix-structure of the Hamiltonian adapted for a quantum well from its pristine bulk form, we refer the reader to one of our recently published works on PbTe~\cite{sengupta2018spin}. The choice of PbTe is largely motivated by its wide applicability in the design of thermoelectric devices~\cite{harman1996high,gelbstein2005high}. Further, since the preceding analysis ignores a reduction of the current via electron-phonon scattering events, we must choose a temperature range that suppresses any lattice vibration; for calculations reported here, the temperature values (in $ K $) lie in the range : $ 50 \leqslant T \leqslant 150 $. Since a purported task of this report is to also compare the quantum of thermally-driven current between parabolic- and linear-dispersion materials, for the latter, we select graphene grown on a substrate that gaps the Dirac cones at the $ K $ and $ K^{'} $ edges of the Brillouin zone. A large number of substrates have been investigated that lower the $ C_{6v} $ point group symmetry of pristine graphene to $ C_{3v} $ introducing a band gap and turning it into a semiconductor; the chief candidates being hexagonal boron nitride (h-BN), silicon carbide (SiC), and oxides such as Al$_{2}$O$_{3}$ and MgO. As a case in point and which serves our goal well by virtue of a more accurate comparison between parabolic and linear materials, graphene grown on Al-terminated Al$_{2}$O$_{3}$ surface is predicted~\cite{skomski2014sublattice} to have a band gap of $ 0.18\, eV $, a number closely aligned to the $ 0.19\,eV $ adopted for representative PbTe. In light of this brief note, we set the linear band gap $\left(\Delta_{L}\right)$ equal to its parabolic counterpart $\left(\Delta = \Delta_{L} = \Delta_{P}\right) $, which is $ 0.19\, eV $. The current that flows in the setup of Fig.~\ref{scfig1} only under a temperature gradient is compared in Fig. for linear and parabolic materials for a series of externally adjusted temperatures $\left(T_{R}\right)$ of the right reservoir. The Fermi level for both class of materials was set to the bottom of the conduction band and an energy window of $ 65\, meV $ was assumed for the numerical integration of Eq.~\ref{lbfcun}. The temperature difference between the reservoir $\left( T_{L} - T_{R}\right) $ was fixed to $ 25\, K $. As for the choice of energy levels, we adopt the following reference levels: The bottom of the conduction band (CB) for the parabolic material is identical to the band gap $\left(\Delta\right)$, and therefore the top of the valence band (VB) is zero. The semi-metal graphene (with both CB and VB equi-energetic at $ 0\, eV $) when gapped by a substrate symmetrically splits the CB and VB bands; the bottom of the CB is located at $ \Delta/2 $ and top of the VB is pushed down to -$\Delta/2 $.

\begin{figure*}
\centering
\includegraphics[width=0.7\textwidth]{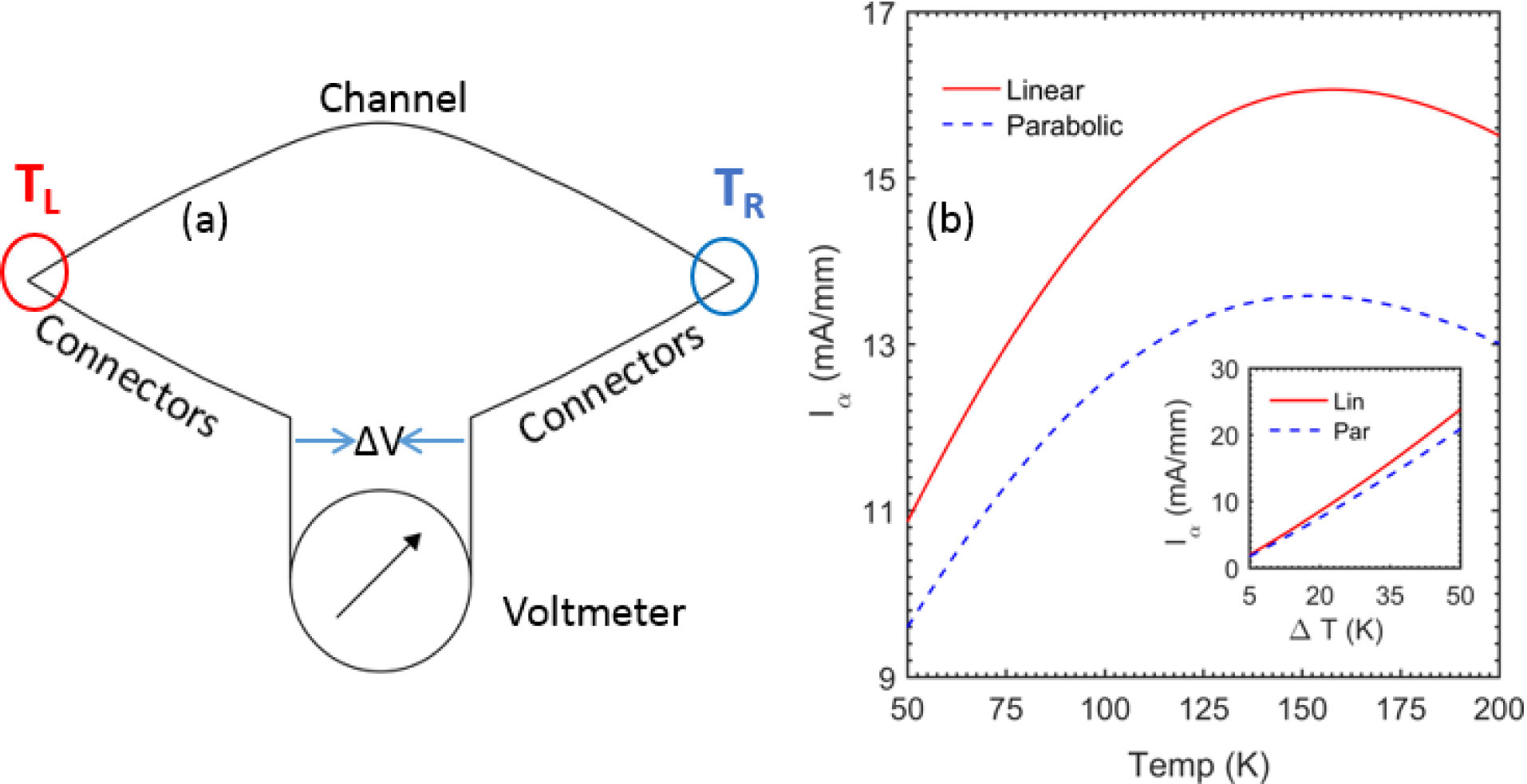}
\caption{The numerically calculated thermal current for a fixed temperature difference $\left(\Delta T\right)$ between the reservoirs (see Fig.~\ref{scfig1}) when linear and parabolic materials are used is shown on the right panel (b). For calculation/numerical parameters described in the text, the linear material offers a greater quantum of thermal current in comparison to the parabolic case, attributable to a higher density of states in the former. Note that the current flows because of a temperature difference induced inequality in the Fermi level of the reservoirs/contacts. The numbers on the \textit{x}-axis indicate the temperature of the right reservoir $\left(T_{R}\right)$; the left reservoir temperature $\left(T_{L}\right)$ for this plot is therefore $ T_{L} = T_{R} + \Delta T $. As further evidence of this point, the inset plot shows current as a function of $\left(\Delta T\right)$ between the reservoirs; a higher difference is reflected in a rising thermal current. Specifically, $ T_{R} = 50\,K $ and $ T_{L} $ as before is higher by $ \Delta T $. Note that no net current actually flows (unless the circuit is completed via connectors) and only an open circuit voltage is measured as shown by the left panel schematic (a). In an open circuit case (vanishing net current), the initial response to $ \Delta T $ is a diffusion current which is exactly counteracted by a internal drift component setup by an electric field oppositely directed to the temperature gradient vector.}
\label{linpar}
\end{figure*}

We pause to point out few key aspects of Fig.~\ref{linpar} before quantitatively analyzing the central theme of this report - the control of thermal currents using a periodic perturbation. First of all, notice that the currents for both linear and parabolic materials acquire a plateau-like profile as the operating temperature increases indicating a drop in the difference of the Fermi level between the reservoirs. It is worthy to bear in mind though that raising the temperature contributes to a re-distribution of the electron population entailing a partial filling of states positioned at higher energies and thus expanding the modes available for current conduction. In principle, the current flow is an interplay of the difference between the Fermi levels of the reservoirs/contacts and the conducting modes in the constricted channel; further, since the overall current flattens out, it is reasonable to state that an increase in the modes notwithstanding, the Fermi level drop between the reservoirs/contacts - which truncates at higher temperature - governs the process. We next note that a temperature difference leads to flow of oppositely directed electron and hole currents; here, the hole contribution though hasn't been explicitly factored into the calculations since the channel material has been chosen to be of \textit{n}-type. The Fermi level lies within the conduction bands, which simply points to a complete filling of the valence states in both the reservoirs and thus a vanishing hole current contribution. Lastly, in graphene-like materials (including cases with a finite gap) which possess particle-hole symmetry, the thermal hole current in a \textit{p}-doped system (the Fermi level aligned to top of the valence band) is exactly opposite to its \textit{n}-type counterpart. The Fermi level for the \textit{n}-type is at the bottom of the conduction band. In conventional parabolic materials, while the hole current is still oppositely directed, an asymmetry vis-\'a-vis the electron current appears as an obvious eventuality of the absence of symmetry between the conduction and valence bands. The hole dispersion (valence bands) is marked by a large warping and significant departures from a `textbook' parabolic model of the conduction states.

\subsection{THz radiation tunable thermopower}
\begin{figure}[t!]
\centering
\includegraphics[scale=0.7]{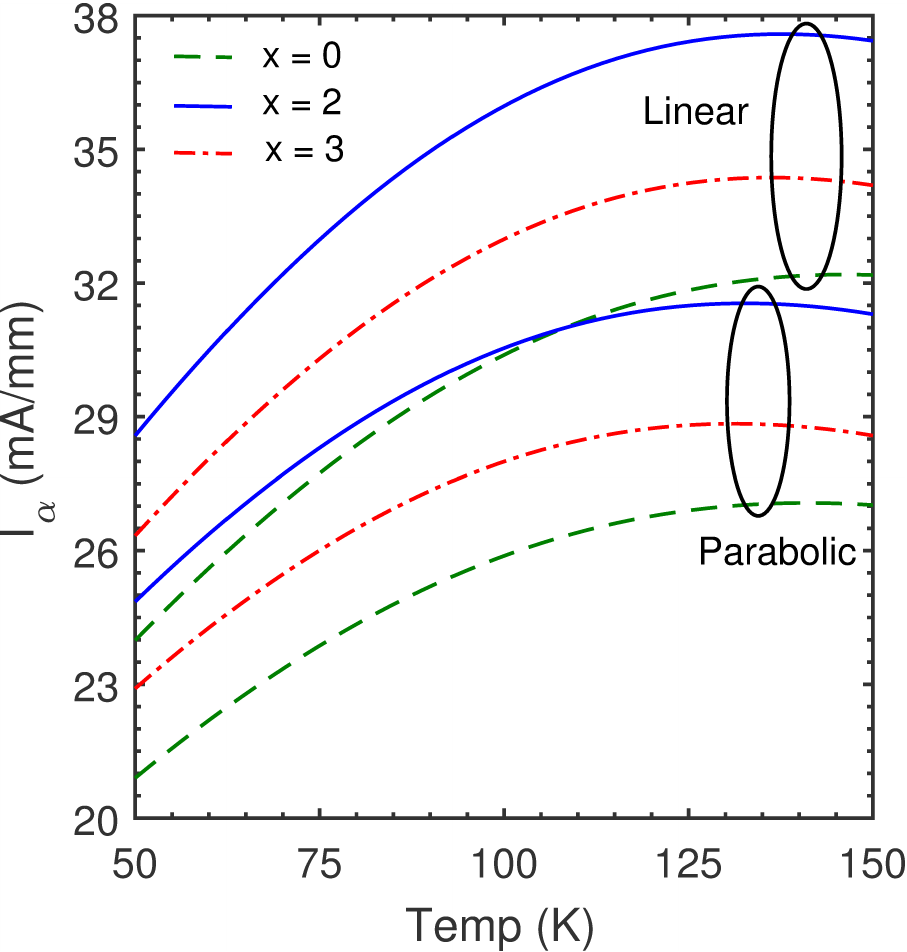}
\caption{The thermopower curves for linear and parabolic materials with an \textit{ac}-source (periodic perturbation) fitted to the left reservoir are plotted using Eq.~\ref{lbfcun}. The frequency of the \textit{ac}-source was set to 1.25 THz. Clearly, for both linear and parabolic materials, the thermopower output is higher when the periodic perturbation induced Floquet states are included in the analysis. The curves marked by $ x = 0 $ denote a thermopower driven solely by the applied temperature gradient. We also show plots for two simulated drive strengths, $ x = 2 $ (solid line) and $ x = 3 $ (broken line); easily, as described in the text, a control of this real-time adjustable parameter can control the flow of the thermal current $\left(I_{\alpha}\right)$ over a range of temperatures. Note that $ x =0 $ (dotted line) represents the case when the periodic perturbation is absent. The temperature as shown on the \textit{x}-axis is that of the right reservoir and of exact description as noted in the caption of Fig.~\ref{linpar} while $ \Delta T $ was set to $ 50\, K $.}
\label{acdrv}
\vspace{-0.53cm}
\end{figure}

The basic character of temperature-driven current curves and the changes thereunto by turning on the periodic perturbation will now be quantitatively examined. We must first, however, select a representative THz frequency suitable for our purpose. The rationale behind the use of THz sources, currently available from a wide array of laser modes (generated via ultra-femtosecond lasers) was briefly touched upon in the introductory section. In calculations here, the frequency $\left(\nu\right)$ is set to 1.25 THz $\left(h\nu \approx 5.0\,meV\right)$, which lies within the 0.3 to 3.0 THz range produced by the popular Auston switches~\cite{nunnally1985photoconductive,auston1975picosecond} based on photo-excitation of carriers into the conduction band of typical semiconductors such as Si, GaAs. The excitation is through short laser pulses. The Floquet states of interest that arise from irradiation with this periodic frequency source by retaining the first two side bands (see Fig.~\ref{bescoff} and related caption and additional discussion in Section~\ref{s3}) are $ \epsilon_{0} + h\nu, \epsilon_{0} + 2h\nu $. Here, $ \epsilon_{0} $ is the base of the conduction band set to the band gap energy $\left(\Delta\right)$ for both linear and parabolic materials. We can now use Eq.~\ref{lbfcun} to establish a thermal current that flows in the channel (Fig.~\ref{scfig1}) with `additional' Floquet states introduced in the left reservoir. While the current curve should exhibit a similar profile as in Fig.~\ref{linpar}, however, we must account for the extra term $ J_{n}^{2}\left(x\right) $ in Eq.~\ref{lbfcun}. This term, which we recall from a preceding discussion, is the probability of the occurrence of additional Floquet states and is closely connected to chosen value of $ x $. For our case, the experimentally-adjustable $ x $ parameter is assigned a pair of values : $ x \in \lbrace 2, 3\rbrace $. The thermal current for the \textit{ac}-driven setup is shown in Fig.~\ref{acdrv}. The material parameters and position of Fermi level are unchanged from those used in preparation of Fig.~\ref{linpar}. The temperature difference between the reservoirs $\left(T_{L} > T_{R}\right)$ was however increased to $ \Delta T = 50\, K $.

The curves (for a pair of values of the drive ratio, $ x $; $ x = 0 $ indicates the absence of any periodic perturbation) in Fig.~\ref{acdrv} easily reveal that a higher current can be pumped for an \textit{ac}-driven setup that introduces Floquet states - an observation simply explained by accounting for the supplementary energy levels that make it possible to inject more thermally activated carriers into the channel. The second straightforward piece of information from the aforementioned figure is a higher quantum of current for a reduced value of $ x $, an observation that follows directly from the nature of the Bessel curves. To understand this further, recall from Section~\ref{s3} - wherein a cutoff to suitably truncate the series expansion of the Bessel's function (Eq.~\ref{besexp}) was determined - that for values of $ x $ significantly above unity $\left(x \gg 1\right)$, the Bessel function places more weight for higher order expansions (defined by a large $ n $) shown by the rightward shifted positive curves in Fig.~\ref{bescoff}. For large $ x $, this manifests as more energetic Floquet bands (the band energies vary as $ \epsilon_{0} + n\hbar\omega $) and with an increased probability $\left(J_{n}^{2}\left(x\right)\right)$ of occurrence. As an aside, $ \sum\limits_{-\infty}^{\infty}J_{n}^{2}\left(x\right) = 1 $, which ensures conservation of probability. While a larger $ J_{n}^{2}\left(x\right) $ adds to the current by virtue of the robustness of the Floquet state, the final outcome is intimately inter-linked to the difference in Fermi levels between the injecting states from the left reservoir and its counterpart in the right maintained at a lower temperature. Indeed, for energies beyond the $ n = 0 $ and $ n = 1 $ mark, the difference in Fermi level between the two reservoirs/contacts rapidly diminishes and higher-placed Floquet energy states negligibly contribute to the thermal current, their higher probability of occurrence notwithstanding. Lastly, a higher current flow is again observed in linear materials vis-\'a-vis those with parabolic bands for reasons that described an identical characteristic of the plots (without an \textit{ac}-perturbation) in Fig.~\ref{linpar}.

The preceding theoretical consideration aside, the rise and fall of the thermal current centered around the drive strength $\left(x\right)$ presents itself as an effective control `tool' to regulate the injection of electrons from the left reservoir to the channel. As a matter of fact, both the drive amplitude $ V_{1} $ and the energy quantum $ h\nu $ can be simultaneously adjusted (via an \textit{ac}-driven gate electrode) altering the ratio $ x = V_{1}/h\nu $ and thus leading to a possible quenching or expansion of the current - suitably illustrated in Fig.~\ref{acdrv}. This mechanism of altering the drive strength is in addition and independent of the temperature gradient (between the reservoirs) assisted modulation. Before closing, a set of comments are in order here: The first remark touches upon the crucial thermodynamic efficiency of the current arrangement that removes heat from the `hot' reservoir and carries it via the production of an electric voltage to the `cold' reservoir. To expound on this point, notice that useful `work' $\left(\mathcal{W}\right)$ in the sense of a Carnot engine cycle is the flow of electric current (a consequence of the developed electric voltage) and the attendant Joule heat when an external load $\left(R\right)$ is attached to the schematic in Fig.~\ref{scfig1}. The efficiency is then simply $ \eta = W/E_{in} $, where $ E_{in} $ is the supplied input energy. For cases where an external periodic perturbation is present, $ E_{in} $ must account for the required energy that drives the \textit{ac}-source along with the heat source to sustain a temperature gradient between the reservoirs. In fact, the whole argument can be given a mathematical structure by noting the expression for efficiency as recorded in an actual engine operating on thermoelectric principles is~\cite{ludovico2016adiabatic}
\begin{equation}
\begin{aligned}
\eta &= \dfrac{T_{L} - T_{R}}{T_{L}},\\
\eta^{'} &= \eta\dfrac{\sqrt{1 + ZT} - 1}{\sqrt{1 + ZT} + 1}.
\label{effcar}
\end{aligned}
\end{equation}
In Eq.~\ref{effcar}, $ \eta^{'} $ is the thermal engine specific efficiency controlled by the figure of merit $ ZT $, a quantity directly linked to the energy levels participating in the thermal process together with electric and thermal conductivity. It therefore suffices to say that the optimization of $ \eta^{'} $ requires a complete analysis of the material properties including a quantitative evaluation of $ E_{in} $ - a task postponed for a later work. 

The second comment is a qualitative statement on the subject of the frequency of the \textit{ac}-source: We explained previously how both $ V_{1} $ and $ h\nu $ through the ratio $ x = V_{1}/h\nu $ can modulate the thermal current; however, it is also important to bear in mind that while a rapidly fluctuating \textit{ac}-signal (high $ \omega = 2\pi\nu $) can adjust $ x $, the system does not fully react to this perturbation, which stated otherwise means the time-average of such an \textit{ac}-field is zero with feeble effects of the \textit{ac}-driving. A long enough settling time $\left(T = 1/\nu\right)$ is therefore a desired prerequisite to observe some of the predicted outcomes of periodically-driven systems. Finally, notice that we did not take into account the heat produced through the \textit{ac}-driving source, which in general can alter the temperature gradient; the heat produced being simply $ \langle V_{1}^{2}\rangle/R $. The angled brackets enclose an average quantity. A natural upshot of this is a realizable thermopower in absence of an external temperature gradient, solely dependent on the periodic driving. Presumtively, such a photo-thermal design (which entails production of thermopower via heating through illumination of the thermoelectric material with a light beam) may lead to a temperature difference and a concomitant production of an electric voltage, which, in addition to material constants is exclusively dependent on the light beam frequency. 

\section{Concluding Remarks}
\label{s4}

We carried out an analytic calculation including quantum effects to estimate the thermopower generation in a channel whose ends are coupled to reservoirs held at different temperatures. The energy levels of the left reservoir were also rearranged by an \textit{ac}-driven source affixed to the gate electrode permitting a more dynamic control of the thermoelectric behaviour. Beyond the results included in this work, we must note that the outlined approach is inadequate insofar as the estimation of thermopower goes when dissipative effects such as impurity scattering, electron-electron interactions, and the more pressing energy broadening that happens at the junction of the reservoirs exert a significant influence on current flow. A fundamental aspect, besides the aforementioned effects not considered in this study, is the tacit assumption that transfer of carriers between reservoirs does not give rise to entropy, a key condition for achieving optimal efficiency $\left(\eta\right) $ in solid-state thermoelectric systems. Further, observe that the use of a periodically driven left reservoir to rearrange the energy levels is one of the many possible configurations that can be conceived to study the problem of energy transfer in a non-equilibrium quantum system; for instance, we can rejig the schematic of Fig.~\ref{scfig1} by moving the \textit{ac}-source from the left reservoir to the channel (while maintaining the temperature gradient between the reservoirs) such that energy states therein are only affected. It then remains to be seen as to what constitutes an ideal design through the prediction of an optimized efficiency $\left(\eta^{'}\right)$ following a similar program that begins by estimating the thermal current in each case. We illustrated this approach for one of the many possible designs (sketched in Fig.~\ref{scfig1}) here.

\begin{figure}[t!]
\centering
\includegraphics[scale=1.1]{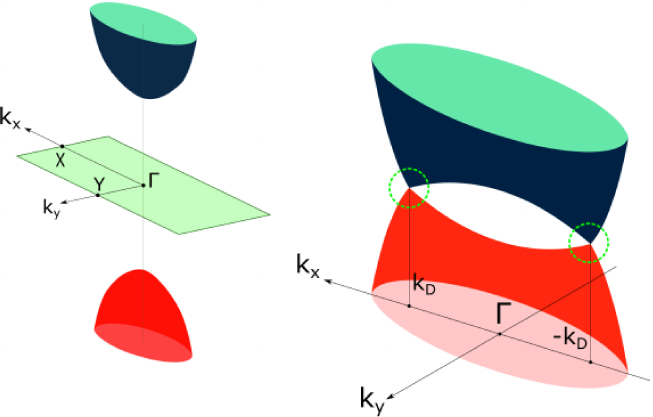}
\vspace{-0.17cm}
\caption{The schematic depiction (left figure) of the dispersion of a semi-Dirac gapped four-layered BP slab. For the semi-Dirac case, the linear dispersion is along the \textit{y}-axis while the \textit{x}-axis hosts a conventional parabolic band. The band gap closing (right figure) happens when BP is doped with potassium (\textit{K}) and the dopant density reaches a threshold value. The band gap closing at non-$\Gamma$ points is shown by dotted circles. The dopant (\textit{K}) induces an electric field which modulates the band gap. As an illustration of a multi-layered structure, four BP sheets stacked together along an out-of-plane axis (inter-layer spacing is $ 5.3\,\AA $) give rise to a semi-Dirac dispersion with a finite gap adjustable through \textit{K}-doping. The single-layer BP is actually a double layered structure and has two P-P bonds. The shorter bond (bond length is $ 2.22\,\AA $) connects the nearest P atoms in the same plane while the longer bond ($ 2.24\,\AA $) connects atoms located in the top and bottom layers of the unit cell.}
\label{dispfig}
\vspace{-0.3cm}
\end{figure}

In this manuscript we have used prototypical linear and parabolic materials to construct thermopower engines; however, the list can be expanded further by the addition of semi-Dirac materials which are essentially a hybrid of the former two classes. A recently discovered semi-Dirac material is layered black phosphorus (BP), a well-known thermoelectric~\cite{ling2015renaissance} with high values of \textit{ZT}; in this particular instance, a stack of BP when doped with potassium~\cite{kim2015observation} gives rise to a parabolic branch along the zigzag (\textit{x}-axis) and linear dispersion in the armchair direction, respectively (Fig.~\ref{dispfig}). The Hamiltonian~\cite{baik2015emergence} is succinctly expressed as $ H_{BP} = \alpha k_{x}^{2} +  \beta k_{y} + \Delta\tau_{z} $, where $ \alpha $ and $ \beta $ are material constants, $ \Delta $ is the band gap, and $ \tau_{z} $ is the sub-lattice degree-of-freedom. The band gap can be closed for a critical dopant concentration. A recently published work~\cite{sengupta2018electrothermal} by us demonstrated a thermopower generator with four-layered \textit{K}-doped BP with semi-Dirac dispersion; the underlying semi-Dirac nature was found to offer superior thermoelectric performance vis-\'a-vis parabolic and linear materials. It is therefore of interest to re-examine the workings and thermodynamic efficiency of a similar BP-based setup in presence of a periodic driving source in tandem with varying potassium doping levels (and a temperature gradient) offering newer insights into thermoelectric behaviour with \textit{ac}-harmonics.

\end{document}